\documentclass[prb,aps,graphics,epsfig,floats,superscriptaddress,preprint]{revtex4}
\hfuzz=\maxdimen
\usepackage{graphicx}
\usepackage{amssymb}
\usepackage{amsmath}
\usepackage{color}

\begin{document}

\title{Spin orbit coupling controlled spin pumping effect}

\author{L. Ma}
\affiliation{Shanghai Key Laboratory of Special Artificial Microstructure Materials and Technology and Pohl Institute of Solid State Physics and School of Physics Science and Engineering, Tongji University, Shanghai 200092, China}
\author{H. A. Zhou}
\affiliation{ The Key Lab for Magnetism and Magnetic Materials of Ministry of Education, Lanzhou University, Lanzhou 730000, People's Republic of China}
\author{L. Wang}
\affiliation{Department of Physics, Beijing Normal University, Beijing 100875, China}
\author{X. L. Fan}
\affiliation{ The Key Lab for Magnetism and Magnetic Materials of Ministry of Education, Lanzhou University, Lanzhou 730000, People's Republic of China}
\author{W. J. Fan}
\affiliation{Shanghai Key Laboratory of Special Artificial Microstructure Materials and Technology and Pohl Institute of Solid State Physics and School of Physics Science and Engineering, Tongji University, Shanghai 200092, China}
\author{D. S. Xue}
\affiliation{ The Key Lab for Magnetism and Magnetic Materials of Ministry of Education, Lanzhou University, Lanzhou 730000, People's Republic of China}
\author{K. Xia}
\affiliation{Department of Physics, Beijing Normal University, Beijing 100875, China}
\author{G. Y. Guo}
\affiliation{Department of Physics, National Taiwan University, Taipei 10617, Taiwan}
\affiliation{Physics Division, National Center for Theoretical Sciences, Hsinchu 30013, Taiwan}
\author{S. M. Zhou}
\affiliation{Shanghai Key Laboratory of Special Artificial Microstructure Materials and Technology and Pohl Institute of Solid State Physics and School of Physics Science and Engineering, Tongji University, Shanghai 200092, China}

\date{\today}
\vspace{5cm}

\begin{abstract}
Effective spin mixing conductance (ESMC) across the nonmagnetic metal (NM)/ferromagnet interface, spin Hall conductivity (SHC) and spin diffusion length (SDL) in the NM layer govern the functionality and performance of pure spin current devices with spin pumping technique. We show that all three parameters can be tuned significantly by the spin orbit coupling (SOC) strength of the NM layer in systems consisting of ferromagnetic insulating Y$_3$Fe$_5$O$_{12}$ layer and metallic Pd$_{1-x}$Pt$_x$ layer. Surprisingly, the ESMC is observed to increase significantly with $x$ changing from 0 to 1.0. The SHC in PdPt alloys, dominated by the intrinsic term, is enhanced notably with increasing $x$. Meanwhile, the SDL is found to decrease when Pd atoms are replaced by heavier Pt atoms, validating the SOC induced spin flip scattering model in polyvalent PdPt alloys. The capabilities of both spin current generation and spin charge conversion are largely heightened via the SOC. These findings highlight the multifold tuning effects of the SOC in developing the new generation of spintronic devices.
\end{abstract}


\maketitle

\begin{figure}[b]
\begin{center}
\includegraphics[width=10cm]{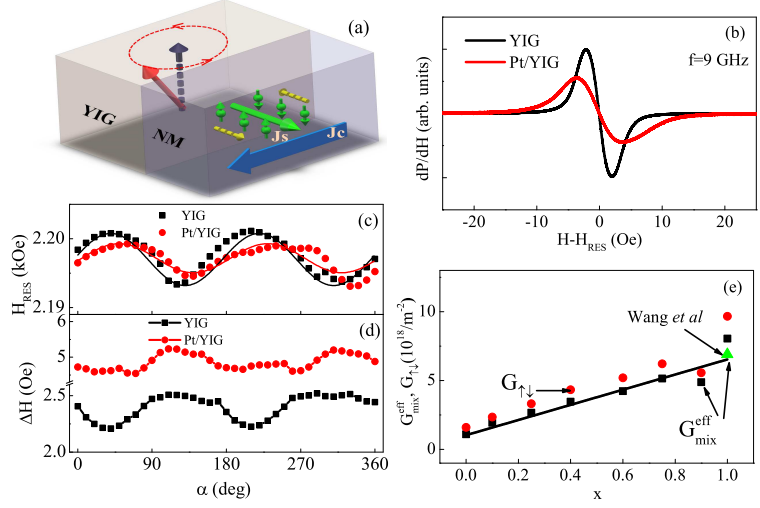}
\caption{Geometry of spin pumping (a). Typical FMR spectra (b), angular dependencies of the resonance field (c) and the resonance linewidth (d) in 80 nm thick YIG and Pt(15 nm)/YIG(80 nm).
 In (e), measured $G_{mix}^{eff}$ (black solid boxes) and SMC $G_{\uparrow\downarrow}$ (red circles) for PdPt(15 nm)/YIG(80 nm), and measured data of Pt/YIG from Wang \textit{et al} (blue triangle)~\cite{29} are given. Solid line in (e) serves a guide to the eye. In (b, c, d, e), $T=300$ K. } \label{Fig1}
\end{center}
\end{figure}

\indent With the prominent advantage of negligible Joule heat, the spin current plays a central role in the next generation of spintronic devices~\cite{Jungwirth2012,4,Miron2011}. Functionality and performance of pure spin current devices strongly depend on the generation and detection approaches of spin current. Among various generation approaches~\cite{2,8,Uchida2008}, the spin pumping has been widely used~\cite{7,8,9,Czeschka2011}, in which the spin current is produced in a heavy non-magnetic metallic (NM) layer when the magnetization precession of the neighboring ferromagnet (FM) layer is excited by the microwave magnetic field, as shown in Fig.~\ref{Fig1}(a). Among a variety of detection techniques~\cite{2,Valenzuela2006}, the conventional electric approach is often employed to probe the spin current via inverse spin Hall effect (ISHE), where the spin current is converted into a charge current.\\

\indent Since their original prediction~\cite{Dyakonov1971,Dyakonov1971a}, the spin Hall effect (SHE) and ISHE have become increasingly important because of their intriguing physics and
great applications in the charge-spin conversion~\cite{1,2,Zhang2000,29,Hoffmann2013,22,8,Valenzuela2006}. It is commonly known that the SHE arises from the spin orbit coupling (SOC)~\cite{Hoffmann2013}. As an outstanding issue, however, the quantitative dependence of the SHE on the SOC strength is still unclear. Moreover, the SHE is known to be contributed by the intrinsic, the skew scattering, and the side-jump terms~\cite{Lowitzer2011,32Morota,39,40,41,Hoffmann2013}. The issue of the SHE scaling law has not been understood yet although many attempts have been made. For instance, the intrinsic term has a dominant contribution to the spin Hall conductivity (SHC) in Ir doped Pt~\cite{Lowitzer2011} whereas the extrinsic skew scattering plays a major role in Ir doped Cu~\cite{39,40,41}. In order to enhance the charge-spin conversion efficiency and to reveal the mechanism of the SHE, it is imperative to study the SHE in non-magnetic alloys in which the SOC strength can be tuned continuously in a wide range. \\

\indent\indent The effective spin mixing conductance (ESMC) across the NM/FM interface has elicited a great deal of attention because it governs the spin pumping efficiency ~\cite{13,14,15,18,Liu2014}, i.e., the density of the dc spin current $\j_{s}$ as follows~\cite{9,18}, $\j_{s}=\frac{\hbar\omega}{4\pi}G_{mix}^{eff}\sin^2\theta$, where $G_{mix}^{eff}$ is the real part of the complex ESMC, the cone angle $\theta$ of the FM magnetization procession is determined by the ferromagnetic resonance (FMR) power absorption, $\omega=2\pi f$ with the radio frequency $f$. It is appealing to reveal the physical mechanism of the ESMC because it is intimately related to the electronic band structure of the NM layer and chemical states on the surface of the FM layer~\cite{13,14}. As well known, $G_{mix}^{eff}$ is identified by the difference in the FMR full width at half maximum (FWHM) between NM/FM and FM, $\Delta H_{NM/FM}-\Delta H_{FM}$ via the following equation~\cite{9}
\begin{equation}
G_{mix}^{eff}=\frac{4\pi\gamma M_st_{FM}}{g\mu_B\omega}(\Delta H_{NM/FM}-\Delta H_{FM}),
\label{pauli2}
\hspace{5.0 cm}
\end{equation}
where $\gamma$ is the gyromagnetic ratio, $g$ the Land\'{e} factor, $t_{FM}$ the thickness of the FM layer, and $M_s$ the magnetization of the FM layer. Since the change in the resonance linewidth of the FM layer is as small as a few oersteds after it is covered by a NM layer~\cite{9,18,13,14,15,Czeschka2011,Liu2014}, much caution must be taken in order to accurately evaluate the value of $G_{mix}^{eff}$. \\

\indent As a critical parameter in spintronic devices, the spin diffusion length (SDL, $\lambda_{sd}$) controls the propagation of the spin current in the NM layer, leading to a strong dependence of the ISHE voltage on the NM layer thickness. The mechanism of the SDL is in sharp debate although it has been studied extensively in both experiments and theory. For example, values of  $\lambda_{sd}$ in Pt measured by various research groups are not consistent~\cite{29zhang,Isasa2015,29,32Morota}, which is suggested to arise from different measurement approaches~\cite{33Chen}. In particular, although the SDL in heavy element NM layers such as Pt is assumed to arise from the spin flip scattering which is in turn caused via the SOC, the dependence of the SDL on the SOC strength is still an open question~\cite{32}. \\

\indent  The objective of this work is to study the ESMC, the SHE, and the SDL as a function of the SOC strength by implementing the Pd$_{1-x}$Pt$_x$ (PdPt)/Y$_3$Fe$_5$O$_{12}$ (YIG) heterostructures, where the SOC can be tuned significantly whereas other physical properties are almost fixed. The SHC and in particular the ESMC are significantly enhanced via changing $x$ from 0 to 1.0. Meanwhile, the SDL decreases with increasing $x$. These phenomena can be attributed to the SOC tuning effects. It will provide novel means to enhance the spin pumping efficiency and to improve the performance of spintronic devices. This work will also be helpful for the community in the newly emerging research field, i.e., the spin-orbitronics~\cite{24}.\\

\indent  First, we study the effect of the SOC on the ESMC. FMR spectra of YIG and Pt/YIG at $f=9.0$ GHz in Fig.~\ref{Fig1}(b) can be described by Lorentz function~\cite{9}. In order to rigorously obtain the ESMC, the in-plane angular dependent FMR spectra of YIG and NM/YIG were measured because the resonance field and the linewidth both depend on the orientation of the in-plane $H$, as shown in Figs.~\ref{Fig1}(c) and ~\ref{Fig1}(d). The angular dependencies of the resonance field in YIG and Pt/YIG can be fitted by considering the in-plane uniaxial anisotropy~\cite{26} and the anisotropy energy is evaluated to be about 340 erg/cm$^3$, much smaller than that of sputtered epitaxial YIG films on Gd$_{3}$Ga$_{5}$O$_{12}$ (GGG) substrates~\cite{28}. The weak magnetic anisotropy indicates the strain relaxation of the YIG films. With the measured frequency dependence of the resonance magnetic field~\cite{26}, $t_{FM}=80$ nm, and $M_S$=136 emu/cm$^3$, the gyromagnetic ratio $\gamma$ and the Land\'{e} factor $g$ are fitted to be 17.57 GHz/kOe and 2.0, respectively. Here, the accuracy of $\Delta H_{NM/FM}$ and $\Delta H_{FM}$ in Eq.~\ref{pauli2} is significantly enhanced by averaging the data points at all orientations. The linewidth enhancement in Pt/YIG arises from the energy transfer from the YIG to the NM layers when the spin angular momentum passes the interface during the spin pumping~\cite{9}. Surprisingly, $G_{mix}^{eff}$ increases with increasing $x$, as shown in Fig.~\ref{Fig1}(e). For the present Pt/YIG, $G_{mix}^{eff}$ is $7.9\times 10^{18}$ m$^{-2}$, close to the results ($6.9\times 10^{18}$ m$^{-2}$) of Wang~\textit{et al.}~\cite{29}. As pointed out by Tserkovnyak and Jiao \textit{et al.}~\cite{Tserkovnyak2002,Jiao2013,ZhangNP2015}, when $\lambda_{sd}$ is significantly smaller than the NM layer thickness, the real part of the ESMC at the NM/FM interface is modified by the NM layer thickness and has a relationship with that of the spin mixing conductance (SMC) $G_{\uparrow\downarrow}$ as follows, $1/G_{mix}^{eff}=1/G_{\uparrow\downarrow}+1/A$ with the parameter $A=\sigma_{NM}/2\lambda_{sd}$ in the unit of $h/e^2$. Accordingly, the real part of the SMC, $G_{\uparrow\downarrow}$ at the interface is achieved to be slightly larger than the $G_{mix}^{eff}$, as shown in Fig.~\ref{Fig1}(e).\\

\indent  It is significant to analyze the physical mechanism for the evolution of the ESMC with the alloy composition. Since the PdPt/YIG samples all have high film quality, as evidenced by the X-ray reflectivity spectra in Fig.S1 in supplemntary materials~\cite{26}, the effect of the microstructure can be excluded. Secondly, with isoelectric Pd and Pt atoms, the number of the channel of PdPt alloys is expected to be independent of the Pt atomic concentration. Clearly, the variation of the SOC with the Pt atomic concentration is uniquely the physical source for the results in Fig.~\ref{Fig1}(e). As well known, for a system without SOC, the $G_{mix}^{eff}$ is governed by the number of channels available in the normal metal, which is hard to be changed. The strong SOC in the NM layer, acting as an effective magnetic field, provides additional channels for the spin loss, leading to the $G_{mix}^{eff}$ enhancement, which will open a new way to increase the spin pumping through the interface. \\



\begin{figure}[b]
\begin{center}
\includegraphics[width=10cm]{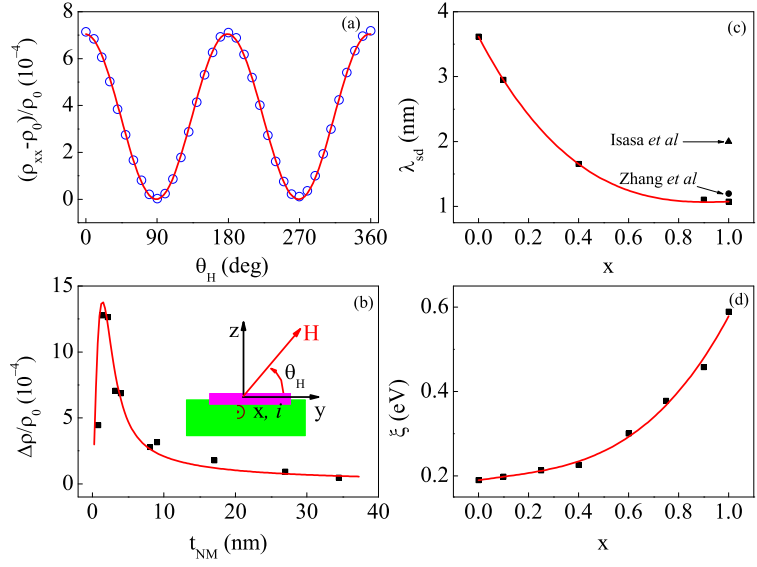}
\caption{For Pt(4 nm)/YIG(80 nm) bilayers, angular dependent SMR (a). For Pt/YIG (80 nm), dependence of the SMR ratio on the Pt layer thickness (b). Measured $\lambda_{sd}$ (c) and calculated $\xi$ (d) in PdPt alloys. Solid lines in (a, b, c, d) refer to the fitted results. In (c, d), the data were fitted in polynomials. In (c), the data of Pt provided by Zhang and Isasa \emph{et al.}~\cite{29zhang,Isasa2015} are also given for comparison. In (a, b), measurements were performed at room temperature.} \label{Fig2}
\end{center}
\end{figure}

\indent We now consider the effect of the SOC on the SDL in the NM layer. With the sensing charge current along the $x$ axis in the film plane as shown in the inset of Fig.~\ref{Fig2}, the NM/YIG exhibits spin Hall magnetoresistance(SMR) effect~\cite{Nakayama2013,Althammer2013} and the sheet longitudinal resistivity obeys the following equation $\rho_{xx}-\rho_0=\Delta\rho m_y^2$, where $m_y$ is the $y$ component of the magnetization unit vector in the YIG layer. As the external magnetic field is rotated in the \emph{yz} plane, the $\rho_{xx}$ changes as a scale of $\cos^2\theta_H$, as observed in Pt/YIG bilayers in Fig.~\ref{Fig2}(a). The $\Delta\rho/\rho_0$ ratio changes non-monotonically with the NM layer thickness, as shown in Fig.~\ref{Fig2}(b). According to the SMR theory, the measured results in Fig.\ref{Fig2}(b)~\cite{26} are fitted and $\lambda_{sd}$ is found to be 1.05 nm for Pt/YIG, which is highly close to the measured results (1.2 nm) of Zhang \emph{et al}~\cite{29zhang}. Clearly, $\lambda_{sd}$ of PdPt alloys at room temperature is found to decrease with increasing $x$, as shown in Fig.~\ref{Fig2}(c). The measured $x$ dependence can be fitted by $\lambda_{sd}=3.616-7.31x+6.85x^2-2.08x^3$, and accordingly the $\lambda_{sd}$ can be obtained for all samples. The SOC strength $\xi$ in PdPt alloys increases as a scale of $Z^{2.56}$ ($Z$=atomic number), as shown by the first-principles calculations in Fig.~\ref{Fig2}(d)~\cite{26}. Apparently, the SDL decreases with increasing $\xi$ as a function of $x$, and in particular $\lambda_{sd}(\mathrm{Pt})/\lambda_{sd}(\mathrm{Pd})\simeq \xi(Pd)/\xi(Pt)$. The results in Figs.~\ref{Fig2}(c) and ~\ref{Fig2}(d) prove well the validity of the spin flip scattering model in PdPt heavy metals~\cite{Marion2014}, in which the SOC induced spin flip scattering rate is proportional to the SOC strength~\cite{32}. Thus, the transport behavior of the pure spin current in PdPt alloys is mainly governed by the Elliot-Yafet spin flip scattering~\cite{Liu2014,32}.\\

\begin{figure}[b]
\begin{center}
\includegraphics[width=10cm]{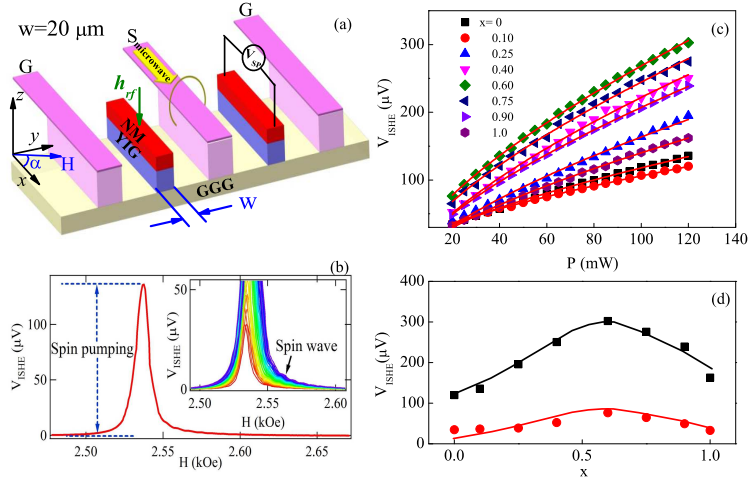}
\caption{(a) Measurement scheme of spin pumping technique. For Pt(15 nm)/YIG(80 nm) sample, $V_{ISHE}$ with the sweeping $H$ at $\alpha=-90$ degrees and an input microwave power $P_{in}=100$ mW (b). For all PdPt(15 nm)/YIG(80 nm) samples, $V_{ISHE}$ as a function of $P_{in}$ (c), and $V_{ISHE}$ versus $x$ at $P_{in}=20$ mW and 120 mW (d). In the inset of (b), $P_{in}$ changes from 20 to 120 mW. Here, $T=300$ K.} \label{Fig3}
\end{center}
\end{figure}

\indent  Then, we investigate the functional dependence of the SHE on the Pt atomic concentration. In experiments of the spin pumping, PdPt/YIG films were patterned into microstrips (2 mm in length, 20 $\mu m$ in width) by using photolithography and ion etching. As shown in Fig.~\ref{Fig3}(a), those strips are placed in the slots between signal (S) and ground (G) of coplanar waveguide (Pt 100 nm) fabricated by sputtering and lift-off techniques. When an alternating current of $f=10$ GHz is applied along the signal line, a radio frequency magnetic field $h_z$ perpendicular to the film plane is induced to trigger the precession of the FM magnetization. $V_{ISHE}$ between both ends of the strip sample along the $x$ axis was detected as a function of $H$. Figure~\ref{Fig3}(b) shows typical $V_{ISHE}$ as a function of $H$ at $\alpha=-90$ degrees, $f=10$ GHz, and $P_{in}=100$ mW for Pt/YIG sample. The maximal voltage is detected at $H_{RES}=¡À2.56$ kOe, the resonance field of the YIG layer. The symmetrical Lorentz line shape (in Fig.~\ref{Fig3}(b)) and the $\sin\alpha$ angular dependent amplitude as shown in supplementary materials~\cite{26} indicate the pure ISHE origin of the resonance voltage~\cite{9}. The inset in Fig.~\ref{Fig3}(b) shows that for Pt/YIG the measured $V_{ISHE}$ increases with increasing $P_{in}$. At high $P_{in}$, the oscillations of the voltage on the high magnetic field side indicate the excitation of spin waves~\cite{Ando2009}. Figure ~\ref{Fig3}(c) shows that for all samples the measured results slightly deviate from the linear dependence, possibly due to the occurrence of nonlinear multimagnon scattering channels~\cite{Jungfl2015}.  Figure~\ref{Fig3}(d) shows that $V_{ISHE}$ changes non-monotonically with $x$ and achieves a maximal value near $x=0.7$. At $P_{in}=120$ mW, the maximal $V_{ISHE}$ is as large as 300 $\mu V$. In the spin pumping technique, $V_{ISHE}$ obeys the following equation~\cite{9}
\begin{equation}
V_{ISHE}=-\frac{eL\omega G_{mix}^{eff}\theta_{SH}\lambda_{sd}\rho}{2\pi t_{NM}}\tanh(\frac{t_{NM}}{2\lambda_{sd}})\theta^2\sin\alpha,
\label{pauli4}
\hspace{5.0 cm}
\end{equation}
with the cone angle of the FM magnetization precession $\theta$, $L=2.0$ mm, $t_{NM}=15$ nm, and the orientation of the magnetization $\alpha=-90$ degrees.

\indent With $V_{ISHE}$ in Fig.~\ref{Fig3}(d) and $\theta$ in supplementary materials~\cite{26} at $P_{in}=20$ mW, and other parameters in Eq.~\ref{pauli4}, $\theta_{SH}$ of PdPt can be deduced for all $x$. In experiments~\cite{26}, $\theta_{SH}$ is found to increase with increasing $x$. For instance, $\theta_{SH}$ is 0.045 and 0.125 for Pd and Pt, respectively. The measured value of the present Pt/YIG is close to the results of 0.12 reported by Zhang \emph{et al}.~\cite{ZhangNP2015}. 
Interestingly, the measured results are also confirmed by the SMR approach~\cite{26}. For example, the value of $\theta_{SH}$ in Pt/YIG is fitted to be 0.120, in agreement with the results by the spin pumping technique. It is noted that the observed variation of $\theta_{SH}$ deviates from the $Z^4$ dependence~\cite{29}, as shown in supplementary materials~\cite{26}.\\

\indent Figures~\ref{Fig4}(a) and~\ref{Fig4}(b) summarize the results of $\sigma_{SH}$ and $\rho$ at 300 K, where $\sigma_{SH}=\theta_{SH}\sigma\hbar/e$ with the electric conductivity $\sigma$ of the NM layer. The measured $\sigma_{SH}$ increases monotonically with increasing $x$. $\sigma_{SH}$ of about 2500 $\hbar/e$ S/cm for Pt is enhanced significantly, compared with that of 665 $\hbar/e$ S/cm for Pd. Interestingly, the measured SHC of Pt/YIG is larger than that of Pd, similar to the calculations~\cite{Guo14}, as shown in Fig.~\ref{Fig4}(a). In contrast, $\rho$ in PdPt layers changes non-monotonically with $x$, as shown in Fig.~\ref{Fig4}(b). The non-monotonic change indicates the random location of the Pd and Pt atoms and the formation of single phase solid solution~\cite{33}.
The measured SHC should in principle consist of the intrinsic and extrinsic terms for $x$ in the region from 0 to 1.0. The extrinsic one, caused by the asymmetric scattering at impurity sites is expected to be negligible near the ending data points~\cite{Lowitzer2011,Guo14} and to become prominent near $x=0.5$. Clearly, for the present PdPt alloys, the intrinsic term plays a dominant role in the measured SHC and the extrinsic term can be neglected, as observed in Ir doped Pt~\cite{Lowitzer2011}. It is the predominant contribution of the intrinsic term that leads to the weak dependence of the spin Hall angle on the NM layer thickness. Indeed, in the analysis of the SDL in PdPt alloys, the measured SMR results in Fig.~\ref{Fig2}(b) can be well fitted assuming the spin Hall angle is independent of the NM layer thickness, hinting the negligible contribution of the surface/interface scattering~\cite{Althammer2013}. It is therefore indicated that the SHC and spin Hall angle in the NM layer can be tuned significantly via the SOC strength. As one of major issues in this field~\cite{Lowitzer2011,32Morota,39,40,41,Hoffmann2013}, the mechanism of the extrinsic SHC needs further investigations. \\

\indent To summarize, a unique heterostructure system consisting of isoelectronic PdPt alloy and YIG is employed to explore the mechanism of the ESMC, the SHE, and the SDL.
Among various physics parameters, only the SOC strength is tuned significantly by changing the alloy composition. The value of $G_{mix}^{eff}$ is rigorously evaluated by measuring the angular dependent resonance linewidth in both YIG and PdPt/YIG. Surprisingly, $G_{mix}^{eff}$ is found to increase when the Pd atoms are replaced by heavier Pt atoms. The $G_{mix}^{eff}$ enhancement is suggested to stem from an increasing SOC strength. It is found that $\sigma_{SH}$ increases with increasing $x$ thanks to the dominant intrinsic contribution. At the same time, the SDL decreases with increasing $x$, indicating the validity of the spin-flip scattering model in PdPt alloys. \\

\indent For PdPt/YIG, the generation efficiency of spin current density $J_s$ is enhanced significantly when $x$ changes from 0 to 1.0. For Pt/YIG, $J_s$ is larger than that of Pd/YIG by a factor of 7. The $J_s$ enhancement will make it easier for the spin current to drive both the magnetization switching and the domain wall motion. Furthermore, the conversion efficiency between the spin current and the charge current in Pt/YIG is larger than that of Pd/YIG by a factor of 2. This offers control over the generation/detection efficiency of spin current and favors to reduce the charge current threshold in spin transfer torque induced magnetization switching. In a word, these findings presented here provide multiple degrees of freedom for improving the functional performance of state-of-the-art spintronic devices and will also provoke further theoretical investigation of the spin dependent transport properties in the NM layers.  \\
\section{Methods}
\indent \textbf{Sample description.} A series of heretostructures consisting of 80 nm thick YIG single-crystal films and polycrystalline PdPt alloy layer were fabricated on (111)-oriented single crystalline Gd$_3$Ga$_5$O$_{12}$ (GGG) substrates via pulsed laser deposition and subsequent DC magnetron sputtering.\\

\indent \textbf{Experimental method.} The film thickness and microstructure were characterized by x-ray reflection (XRR) and x-ray diffraction (XRD). Magnetization hysteresis loops were measured by physical property measurement system (PPMS). The ISHE voltage $V_{ISHE}$ was detected by spin pumping technique, in combination of the FMR technique. The resonance linewidths of YIG and PdPt/YIG were measured by the FMR technique with in-plane $H$. The SDL was measured by the spin Hall magnetoresistance (SMR) technique~\cite{Althammer2013}. Details of fabrication and measurements are described in supplementary materials~\cite{26}.\\

\indent \textbf{{\it Ab initio} calculations for SOC.} Details of calculations of the SOC strength are described in supplementary materials~\cite{26}.\\ 

\section{Acknowledgement}
\indent This work was supported by the State Key Project of Fundamental Research Grant No. 2015CB921501, the National Science Foundation of China Grant Nos. 11374227, 51331004, 51171129, and 51201114, Shanghai Science and Technology Committee Nos. 0252nm004, 13XD1403700, and 13520722700. \\
\section{Author contributions}
S. M. Z. and X. L. F. designed the experiments. L. M. fabricated the samples. L. M. and H. A. Z. carried out the measurements. G. Y. G. performed the theory calculations. L. M., H. A. Z., X. L. F., W. J. F., D. S. X., L. W., and K. X. analyzed the data. L. M. and S. M. Z. wrote the paper and all the co-authors comment on it.

\section{Additional information}
Supplementary information is available in the online version of the paper. Reprints and permissions information is available online at www.nature.com/reprints.
Correspondence and requests for materials should be addressed to S. M. Z. and X. L. F.

\section{ Competing financial interests}
\indent The authors declare no competing financial interests.

\begin{figure}[b]
\begin{center}
\includegraphics[width=10cm]{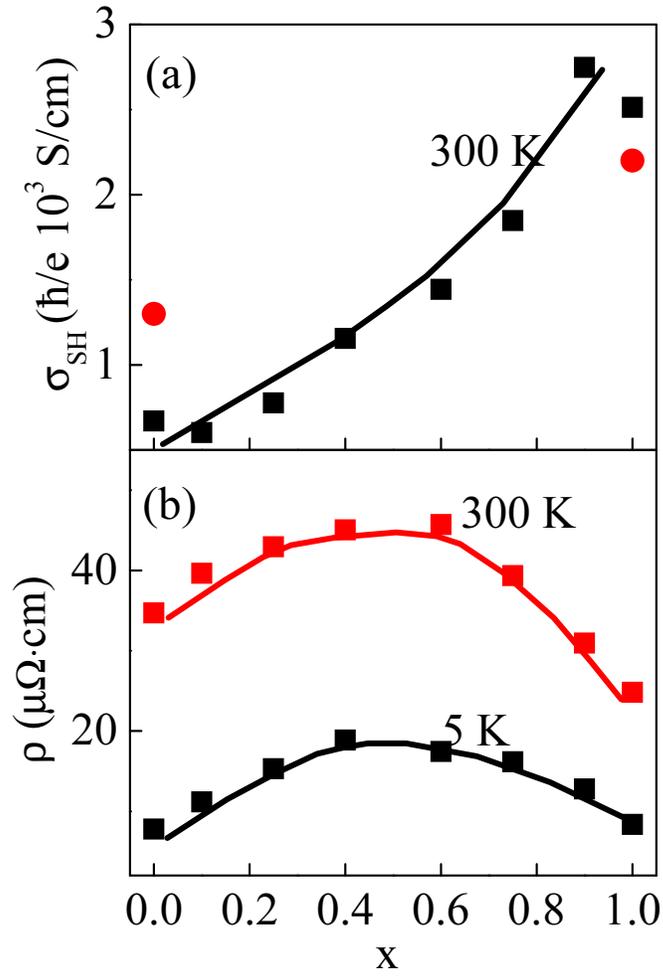}
\caption{For PdPt (15 nm)/YIG (80 nm), measured $\sigma_{SH}$ at 300 K (a) and resistivity at 5 K and 300 K (b) as a function of $x$. In (a) the intrinsic SHC of Pt and Pd are also given from the first-principles relativistic band model (red, solid circles)~\cite{26}. In (a, b), solid lines serve a guide to the eye.} \label{Fig4}
\end{center}
\end{figure}

\end{document}